\documentclass[journal]{IEEEtran}
\usepackage{bm}
\usepackage{amssymb}
\usepackage{amsfonts}
\usepackage{amsmath}
\usepackage{algorithm}
\usepackage{balance}
\usepackage{graphicx,subfig,cite,multicol,multirow,diagbox,booktabs,array}
\usepackage{epstopdf}

\usepackage{color}
\usepackage{float}
\usepackage{stfloats}
\usepackage{makecell}
\ifCLASSINFOpdf
\else
\fi
\begin{document}
\title{STAR-RIS-UAV Aided Coordinated Multipoint Cellular System for Multi-user Networks}

\author{Baihua Shi, Yang Wang, Danqi Li, Wenlong Cai, Jinyong Lin, Shuo Zhang, Weiping Shi, Shihao Yan, and Feng Shu
\thanks{This work was supported in part by the National Natural Science Foundation of China (Nos.U22A2002, 61972093 and 62071234), Hainan Province Science and Technology Special Fund (ZDKJ2021022), and the Scientific Research Fund Project of Hainan University under Grant KYQD(ZR)-21008.}
\thanks{B. Shi, Y. Wang, D. Li, and W. Shi are with the School of Electronic and Optical Engineering, Nanjing University of Science and Technology, Nanjing 210094, China.}
\thanks{F. Shu is with the School of Information and Communication Engineering, Hainan University, Haikou, 570228, China and also with the School of Electronic and Optical Engineering, Nanjing University of Science and Technology, Nanjing, 210094, China. (e-mail: shufeng0101@163.com)}
\thanks{W. Cai, J. Lin, and S. Zhang are with the National Key Laboratory of Science and Technology on Aerospace Intelligence Control, Beijing Aerospace Automatic Control Institute, Beijing 100854, China.}
\thanks{S. Yan is with the School of Science and Security Research Institute, Edith Cowan University, Perth, WA 6027, Australia.}
}
\maketitle

\begin{abstract}
Different with conventional reconfigurable intelligent surface (RIS), simultaneous transmitting and reflecting RIS (STAR-RIS) can reflect and transmit the signals to the receiver. In this paper, to serve more ground users and increase the deployment flexibility, we investigate an unmanned aerial vehicle equipped with a STAR-RIS (STAR-RIS-UAV) aided wireless communications for multi-user networks. Energy splitting (ES) and mode switching (MS) protocols are considered to control the reflection and transmission coefficients of STAR-RIS elements. To maximize the sum rate of the STAR-RIS-UAV aided coordinated multipoint cellular system for multi-user networks, the corresponding beamforming vectors as well as transmitted and reflected coefficients matrices are optimized. Specifically, instead of adopting the alternating optimization, we design an iteration method to optimize all variables for both ES and MS protocols at the same time. Simulation results reveal that STAR-RIS-UAV aided wireless communication system has a much higher sum rate than the system with conventional RIS or without RIS. Furthermore, the proposed structure is more flexible than a fixed STAR-RIS and could greatly promote the sum rate.
\end{abstract}


\IEEEpeerreviewmaketitle

\section{Introduction}
Reconfigurable intelligent surface (RIS) has attracted the attention of many researchers from both the academic and industry field since it emerged. RIS could solve many existed problems (e.g.,High system energy consumption, high hardware costs) in wireless communications by reflecting and controlling the phase of the incident signals \cite{huangRIS2019twc,wuRIS2020cm,liuRIS2021survey}. RIS has been seen as a key technology of the sixth generation communication network (6G). 
And, it has been adopted in many wireless applications, such as energy harvesting system \cite{shiRIS2019cl,panRIS2020jsac,WuQos,WU_SWIPT_IRS}, physical layer security \cite{shenRIS2019cl,cuiRIS2019wcl,shiRIS2022tcom}, non-orthogonal multiple access (NOMA) \cite{yueRISNOMA2023twc,KhaleelRIS2022jstsp,wuRIS2021cl}, convert wireless communications \cite{zhouRIS2022twc}, massive multiple-input multiple-output (MIMO) \cite{zhiRIS2023tit,PAN_IRS_MIMO}, ultra-reliability low-latency communications (URLLC) \cite{renURLLC2022tcom} and so on. Different with the relay station, RIS is a passive device. This means that it consumes little energy and is suitable to be used in unmanned aerial vehicles (UAVs) network \cite{liRIS2020wcl,muRIS2021jsac,renUAVRIS2023iot}.

RIS is a two-dimensional (2D) surface that consisted of a controller and many low-power passive elements. The controller is able to adjust the phase of the incident signal. Then, the element will reflect the signal to the receiver. Thus, RIS is similar to the relay. However, RIS has much lower cost and energy consumption. Given the many advantages of RIS, he has been studied in many literature \cite{wuRIS2019twc,hongRIS2020tcom,shiRIS2021cl,shuRIS2021tcom,dongRIS2022jcn}.
In \cite{wuRIS2019twc}, authors  designed the beamforming for a RIS assisted wireless network and proved that the performance of all users could be greatly improved with the help of RIS. 
Artificial noise (AN) and RIS assisted  secure wireless communications were investigated in \cite{hongRIS2020tcom}. In addition, a block coordinate descent and majorization-minimization based method was proposed to optimize the phase shift of RIS and AN covariance matrix.
Physical layer security and energy harvesting were considered at the same time in \cite{shiRIS2021cl}. And, a novel transmit beamforming method was also proposed to meet the above two requirements.
In \cite{shuRIS2021tcom}, RIS was also used to increase the secrecy rate in directional modulation system. 
In order to decrease more the circuit cost, authors analyzed the performance of discrete-phase-shifter in RIS aided system in \cite{dongRIS2022jcn}.

However, RIS is unhelpful when the user is behind the surface. Owing to the rapid development of material, simultaneous transmitting and reflecting RIS (STAR-RIS) was proposed in recent years \cite{niuSTARRIS2022tvt}.  Since the incident signal could be reflected or refracted by the element of STAR-RIS by adjusting the transmitted and reflected coefficients (TARCs) of elements, the STAR-RIS is able to serve users whether they are on the front or back of the STAR-RIS. There are three operation protocols that can control the transmitted and reflected signals: energy splitting (ES), mode switching (MS), and time switching (TS). 
Thus, many works have been donated to STAR-RIS in recent years \cite{niuSTARRIS2022tvt,muSTARRIS2022twc,yueSTARRIS2023twc,gaoSTARRIS2023tcom,zhaiSTARRIS2023tcom}.
In \cite{niuSTARRIS2022tvt}, performance of three protocols were analyzed. Authors proved that STAR-RIS could achieve better sum rate than conventional RIS. More importantly, TS and ES has the best performance in unicast communication and broadcast communication, respectively. In \cite{yueSTARRIS2023twc}, to meet more stringent quality-of-service (QoS) requirements, authors considered STAR-RIS aided downlink NOMA networks. The joint optimization for the location and beamforming of STAR-RIS was proposed in \cite{gaoSTARRIS2023tcom}. Furthermore, this paper revealed that the optimized position of STAR-RIS could significantly enhance the sum rate.

Recently, to  suppress interference between adjacent  cellulars, coordinated multipoint (CoMP) was proposed \cite{IrmerCoMP2011cm}. And, some RIS-aided CoMP works were conducted \cite{huaRISCoMP2021tcom,xieRISmc2021twc}. However, to the best of our knowledge, STAR-RIS aided CoMP has not been investigated.
Furthermore, the configuration of users is not fixed and it is hard to set all users on the one side of the RIS. Therefore, we integrate the STAR-RIS  into a CoMP system to further improve system performance. In addition, when STAR-RIS is deployed on the ground, the line-of-sight (LOS) part of channel is usually missed due to obstacles. Thus, STAR-RIS is attached to a UAV and the UAV hovers over the preset position. This ensure that the LOS is  existed and the deployment location of STAR-RIS is more flexible.
The main contributions of this paper are summarized as follows:
\begin{enumerate}
	\item A UAV equipped with a STAR-RIS (STAR-RIS-UAV) aided CoMP wireless communications system is proposed in this paper. And, ES and MS protocols are considered in the system. To achieve high performance, the STAR-RIS-UAV is deployed at the middle of two base station (BS). Then, we aim to optimize the beamforming vector and TARCs matrices to maximize the system sum rate. In addition, the transmit power for BS and the QoS of users are considered as constrains.
	\item For ES protocol, different with the conventional alternate optimization that optimize one variable every time, the proposed method could optimize all variables in iterations. Successive convex approximation (SCA) and penalty function are adopted to convert the non-convex problem into a convex problem. Then, the optimized results will be achieved by updating the penalty factor in iterations.
	\item The processing of MS protocol is similar to the ES protocols. The key is to deal with a binary constraint. We replace this  binary constraint with a new penalty function. When the penalty factor increase to infinity, the modified problem is equal to the origin.
	\item We evaluate the performance by comparing the system sum rate of proposed methods with three schemes: no RIS, conventional RIS and uniform energy splitting. Simulation results demonstrate that the proposed STAR-RIS-UAV has the highest system sum rate. And, the performance of ES protocol is better than the MS protocol.
\end{enumerate}
\emph{Notations:} Throughout the paper, vectors and matrices are respectively denoted by $\mathbf{x}$ and $\mathbf{X}$ in bold typeface, while normal typeface is used to represent scalars, such as $x$. Signs $(\cdot)^T$, $(\cdot)^H$, $|\cdot|$ and $\|\cdot\|$ represent transpose, conjugate transpose, modulus and norm, respectively. $\mathbf{I}_M$ denotes the $M\times M$ identity matrix. Furthermore, $\mathbb{E}[\cdot]$ represents the expectation operator, and $\mathbf{x}\sim \mathcal{CN}(\mathbf{m},\mathbf{R})$ denotes a circularly symmetric complex Gaussian stochastic vector with mean vector $\mathbf{m}$ and covariance matrix $\mathbf{R}$.
$\mathrm{diag}(\mathbf{x})$ denotes a diagonal matrix composed of $\mathbf{x}$. $\mathrm{Tr}(\cdot)$ is the matrix trace operation. 

The rest of this paper is organized as follows. The system model of the STAR-RIS-UAV aided CoMP cellular system for multi-user networks is presented in Section \ref{sec_sys}. Then, two penalty function based iterative methods are proposed in Section \ref{sec_met}. The computational complexity is investigated in Section \ref{sec_com}. In Section \ref{sec_simu}, simulation and numerical results are provided to analyze the performance and convergence of proposed methods. Finally, we come to the conclusion in Section \ref{sec_con}.
\section{System Model}\label{sec_sys}
\begin{figure}[H]
	\includegraphics[width=0.48\textwidth]{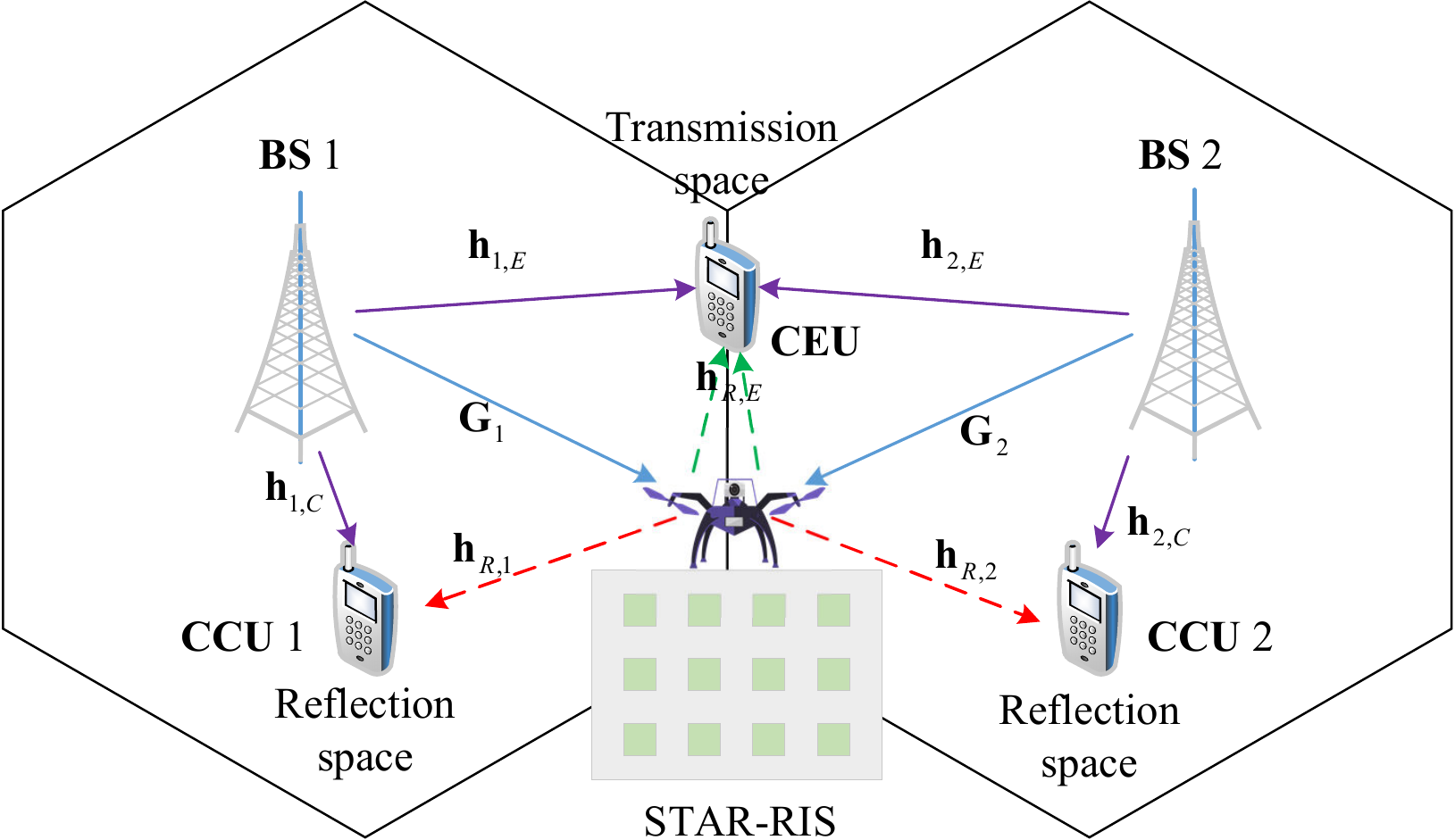}
	\caption{System model.\label{fig_sys}}
\end{figure}   
As shown in Figure \ref{fig_sys}, we consider a STAR-RIS-UAV aided CoMP cellular system that consists of two adjacent cells. Two multiple-antenna BSs are employed for cooperatively supporting a cell edge user (CEU). Additionally, each BS serves an additional cell center user (CCU). 
The STAR-RIS-UAV is hovered at the middle of the two cells to divide the whole system into two spaces, and the CEU is located in the transmission half-space, while the CCU is located in the reflection half-space.
The BS is equipped with $M$ antennas, and the STAR-RIS has $N$ elements. CEU and CCU are both single-antenna users. We denote the channels from the BS to the STAR-RIS, the BS to the CEU, the BS to the CCU, the STAR-RIS to the CEU and the STAR-RIS to the CCU as $\mathbf{G}_{k}(k=1,2)\in\mathbb{C}^{N\times M}$, $\mathbf{h}_{k,E}^H\in\mathbb{C}^{1\times M}$, $\mathbf{h}_{k,C}^H\in\mathbb{C}^{1\times M}$, $\mathbf{h}_{R,E}^H\in\mathbb{C}^{1\times N}$ and $\mathbf{h}_{R,k}^H\in\mathbb{C}^{1\times N}$. In addition, we assume that all the CSIs are perfectly known at the BS.

For ES protocol, all elements of the STAR-RIS operate in the two modes, and the energy of the signal incident on each element is usually split into the transmitted signal energy and the reflected signal energy with an energy splitting ratio of $\beta_{n}^{t}:\beta_{n}^{r}$. Thus, the TARC matrices of the STAR-RIS are respectively given by
\begin{equation}\label{Theta_t}
	\bm{\Theta}_t^{\mathrm{ES}} =\mathrm{diag} \left(\sqrt{\beta_{1}^t}e^{j\theta_{1}^{t}}, \sqrt{\beta_{2}^t}e^{j\theta_{2}^{t}},\cdots,\sqrt{\beta_{N}^t}e^{j\theta_{N}^{t}}\right)
\end{equation}
and
\begin{equation}\label{Theta_r}
	\bm{\Theta}_r^{\mathrm{ES}} =\mathrm{diag} \left(\sqrt{\beta_{1}^r}e^{j\theta_{1}^{r}}, \sqrt{\beta_{2}^t}e^{j\theta_{2}^{r}},\cdots,\sqrt{\beta_{N}^t}e^{j\theta_{N}^{r}}\right),
\end{equation}
where $\beta_{n}^{t},\beta_{n}^{r}\in[0,1]$, $\beta_{n}^{t}+\beta_{n}^{r}=1$, and $\theta_{n}^{t},\theta_{n}^{r}\in[0,2\pi)$, $\forall n\in N$. Since the TARCs of each element can be optimized, a high flexibility of communication system design can be realized.

For MS protocol, all elements of the STAR-RIS are divided into two groups. One group contains $N_t$ elements, which work in the transmission mode, and the other group contains $N_r$ elements working in the reflection mode, where $N_t+N_r=N$. In this case, the TARC matrices of the STAR-RIS are given by 
\begin{equation}
	\bm{\Theta}_{t}^{\rm MS}={\rm diag}\left(\sqrt{\beta_{1}^{t}}e^{j\theta_{1}^{t}},\sqrt{\beta_{2}^{t}}e^{j\theta_{2}^{t}},\cdots,\sqrt{\beta_{N}^{t}}e^{j\theta_{N}^{t}}\right)
\end{equation}
and
\begin{equation}
	\bm{\Theta}_{r}^{\rm MS}={\rm diag}\left(\sqrt{\beta_{1}^{r}}e^{j\theta_{1}^{r}},\sqrt{\beta_{2}^{r}}e^{j\theta_{2}^{r}},\cdots,\sqrt{\beta_{N}^{r}}e^{j\theta_{N}^{r}}\right),
\end{equation}
respectively, where 
$\beta_{n}^{t},\beta_{n}^{r}\in\{0,1\}$, $\beta_{n}^{t}+\beta_{n}^{r}=1$, and $\theta_{n}^{t},\theta_{n}^{r}\in[0,2\pi)$, $\forall n\in N$. MS protocol can be seen as a special case of ES protocol, where the amplitude coefficients of transmission and reflection are limited to 0 or 1. MS typically cannot achieve the same reflection beamforming gains as ES, because only a fraction of the elements are selected for transmission and reflection. However, MS is still attractive in practice, since this "on-off" type operation is easier to achieve than ES.

Let $\mathbf{w}_{k,E}$ and $\mathbf{w}_{k,C}$ denote the beamforming vector for the CEU and CCU, $s_{E}$ and $s_{k,C}$ denote the information symbol transmitted to the CEU and CCU, respectively. Without loss of generality, it is assumed that $\mathbb{E}\{|s_{E}|^2\}=1$ and $\mathbb{E}\{|s_{k,C}|^2\}=1,\forall k$. Then, the received signals at the CEU and CCU are respectively given by
\begin{equation}
	y_{E}=\sum_{k=1}^{2}(\mathbf{h}_{k,E}^{H}+\mathbf{h}_{R,E}^{H}\bm{\Theta}_{t}^{\mathrm{ES/MS}}\mathbf{G}_{k})(\mathbf{w}_{k,E}s_{E}+\mathbf{w}_{k,C}s_{k,C})+n_{E},
\end{equation}
\begin{equation}
	y_{k,C}=(\mathbf{h}_{k,C}^{H}+\mathbf{h}_{R,k}^{H}\bm{\Theta}_{r}^{\mathrm{ES/MS}}\mathbf{G}_{k})(\mathbf{w}_{k,E}s_{E}+\mathbf{w}_{k,C}s_{k,C})+n_{k,C},
\end{equation}
where $n_{E}\sim\mathcal{CN}(0,\sigma_{E}^2)$ and $n_{k,C}\sim\mathcal{CN}(0,\sigma_{k,C}^2)$ are the additive white Gaussian noises at the CEU and CCU, respectively. Let us define 
\begin{align}
	&\mathbf v_{l}=[\sqrt{\beta_{1}^{l}}e^{j\theta_{1}^{l}},\sqrt{\beta_{2}^{l}}e^{j\theta_{2}^{l}},\cdots,\sqrt{\beta_{N}^{l}}e^{j\theta_{N}^{l}}]^H\in\mathbb{C}^{N\times1},\nonumber\\
	&~~~l\in{t,r}, \mathbf{u}_{l}=[\mathbf{v}_{l};1] \nonumber\\
	&\mathbf {F}_{R,k}={\rm diag}\{\mathbf{h}_{R,k}^H\},\mathbf{G}_k\in\mathbb{C}^{N\times M}, \mathbf{F}_{k}=[\mathbf{F}_{R,k};\mathbf{h}_{k,C}^H], \nonumber\\
	&\mathbf {H}_{R,k}={\rm diag}\{\mathbf{h}_{R,E}^H\},\mathbf{G}_k\in\mathbb{C}^{N\times M}, \mathbf{H}_{R}=[\mathbf{H}_{R,1},\mathbf{H}_{R,2}], \nonumber\\
	& \mathbf{h}_{E}=[\mathbf{h}_{1,E}^H,\mathbf{h}_{2,E}^H]^H, \mathbf{H}=[\mathbf{H}_{R};\mathbf{h}_{E}^H],\nonumber\\ &\mathbf{w}_{C}=[\mathbf{w}_{1,C}^{H},\mathbf{w}_{2,C}^H]^H, \mathbf{w}_{E}=[\mathbf{w}_{1,E}^{H},\mathbf{w}_{2,E}^H]^H
\end{align}
for notational simplicity. Therefore, the achievable rate at the CEU and CCU for ES and MS are given by
\begin{equation}\label{R_E}
	R_{E}=\log_2(1+\frac{|\mathbf{u}_{t}^H\mathbf{H}\mathbf{w}_{E}|^2}{|\mathbf{u}_{t}^H\mathbf{H}\mathbf{w}_{C}|^2+\sigma_{E}^2}),
\end{equation}
\begin{equation}\label{R_kc}
	R_{k,C}=\log_2(1+\frac{|\mathbf{u}_{r}^H\mathbf{F}_{k}\mathbf{w}_{k,C}|^2}{|\mathbf{u}_{r}^H\mathbf{F}_{k}\mathbf{w}_{k,E}|^2+\sigma_{k,C}^2}),
\end{equation}
respectively. In this paper, we aim to maximize the sum rate of the STAR-RIS aided CoMP network by jointly optimizing the beamforming vectors and the TARCs, subject to the transmit power constraints for the BSs, and the QoS constraint of the users.

\section{Proposed Iterative Algorithm by Employing Penalty Function}\label{sec_met}
In this section, we propose two penalty-based iterative algorithms to solve the optimization problem for ES and MS protocols. Instead of employing the alternating optimization, we aim to optimize all optimization variables concurrently in each iteration. Then, this algorithm is further extended to solve the problem for MS.

\subsection{Joint Design of Beamforming Vector and STAR-RIS Coefficient Matrix for ES Protocol}
Specifically, for the ES protocol, the optimization problem can be formulated as
\begin{subequations}\label{p1}
	\begin{align}
		&\max\limits_{\mathbf{w}_{k,C},\mathbf{w}_{k,E},\mathbf{u}_{l}}\enspace\sum_{k=1}^{2}R_{k,C}+R_{E}\\
		\text {s.t.}\enspace&\enspace||\mathbf{w}_{k,C}||^2+||\mathbf{w}_{k,E}||^2\leq P,\enspace\forall k\\
		&\enspace R_{k,C}\geq \bar{R}_{k,C},\enspace\forall k\\
		&\enspace R_{E}\geq \bar{R}_{E}\\
		&\enspace[\mathbf{u}_{l}]_{n}=\sqrt{\beta_{n}^{l}}e^{j\theta_{n}^{l}},\theta_{n}^{l}\in[0,2\pi),\enspace\forall l,n\\
		&\enspace[\mathbf{u}_{l}]_{N+1}=1\\
		&\enspace\beta_{n}^{l}\in[0,1],\sum\nolimits_{l}\beta_{n}^{l}=1,\enspace\forall l,n
	\end{align}
\end{subequations}
where $P$ denotes the maximum transmit powers for each BS, $\bar{R}_{E}$ and $\bar{R}_{k,C}$ denote the minimum rate requirement of the CEU and CCU, respectively. For the MS protocol, It is easily known that we can reformulate by replacing $\beta_{n}^{l}\in[0,1]$ with $\beta_{n}^{l}\in{[0,1]}$. We note that it is difficult to obtain the globally optimal solution directly for the non-convex problem in. Specifically, the constraints (\ref{p1}c) and (\ref{p1}d) are non-convex, due to the coupled optimization variables. Furthermore, for the conventional reflecting-only RIS, only one coefficient needs to be designed, while STAR-RIS require the jointly optimization of the TARC, which aggravate the difficulty for solving the optimization problem.

Firstly, we deal with the issue of coupled optimization variables. Let us define $\mathbf{W}_{E}=\mathbf{w}_{E}\mathbf{w}_{E}^H$, $\mathbf{W}_{C}=\mathbf{w}_{C}\mathbf{w}_{C}^H$, $\mathbf{W}_{k,E}=\mathbf{w}_{k,E}\mathbf{w}_{k,E}^H$, $\mathbf{W}_{k,C}=\mathbf{w}_{k,C}\mathbf{w}_{k,C}^H$. Then, the SINR expressions in (\ref{R_E}) and (\ref{R_kc}) can be equivalently rewritten as
\begin{equation}\label{Gamma_E}
	\Gamma_E=\frac{{\rm Tr}(\mathbf{U}_t\mathbf{H}\mathbf{W}_E\mathbf{H}^H)}{{\rm Tr}(\mathbf{U}_t\mathbf{H}\mathbf{W}_C\mathbf{H}^H)+ \sigma_E^2}
\end{equation}
\begin{equation}\label{Gamma_kc}
	\Gamma_{k,C}=\frac{{\rm Tr}(\mathbf{U}_r\mathbf{F}_k\mathbf{W}_{k,C}\mathbf{F}_k^H)}{{\rm Tr}(\mathbf{U}\mathbf{F}_k\mathbf{W}_{k,E}\mathbf{F}_k^H)+\sigma_{k,C}^2}
\end{equation}
By employing (\ref{Gamma_E}) and (\ref{Gamma_kc}), the optimization problem for ES can be reformulated as
\begin{subequations}\label{p2}
	\begin{align}
		&\min\limits_{\mathbf{W}_{k,C},\mathbf{W}_{k,E},\atop\mathbf{W}_{C},\mathbf{W}_{E},\mathbf{U}_{l}}\enspace-\sum_{k=1}^{2}\log_2(1+\Gamma_{k,C})-\log_2(1+\Gamma_{E})\\
		\text {s.t.}\enspace&{\rm Tr}(\mathbf{W}_{k,C})+{\rm Tr}(\mathbf{W}_{k,E})\leq P,\enspace\forall k\\
		&\gamma_{k,C}{\rm Tr}(\mathbf{U}_{r}\mathbf{F}_{k}\mathbf{W}_{k,E}\mathbf{F}_{k}^H)+\gamma_{k,C}\sigma_{k,C}^2\notag\\
		&-{\rm Tr}(\mathbf{U}_{r}\mathbf{F}_{k}\mathbf{W}_{k,C}\mathbf{F}_{k}^H)\leq 0,\enspace\forall k\\
		&\gamma_{E}{\rm Tr}(\mathbf{U}_{t}\mathbf{H}\mathbf{W}_{C}\mathbf{H}^H)+\gamma_{E}\sigma_{E}^2\notag\\
		&-{\rm Tr}(\mathbf{U}_{t}\mathbf{H}\mathbf{W}_{E}\mathbf{H}^H)\leq 0,\\
		&0\leq [\mathbf{U}_{l}]_{n,n}\leq 1,[\mathbf{U}_{t}]_{n,n}+[\mathbf{U}_{r}]_{n,n}=1,\enspace\forall l,n\\
		&[\mathbf{U}_{t}]_{N+1,N+1}=[\mathbf{U}_{r}]_{N+1,N+1}=1\\
		&\mathbf{W}_{C},\mathbf{W}_{E}\succeq0,\mathbf{W}_{k,C},\mathbf{W}_{k,E}\succeq0,\forall k,\\
		&\mathbf{U}_{l}\succeq0,\forall l\\
		&{\rm rank}(\mathbf{W}_{C})=1, {\rm rank}(\mathbf{W}_{C})=1,\notag\\
		&{\rm rank}(\mathbf{W}_{k,C})=1,{\rm rank}(\mathbf{W}_{k,E})=1,\forall k\\
		&{\rm rank}(\mathbf{U}_{l})=1,\forall l
	\end{align}
\end{subequations}
where $\gamma_{E}=2^{\bar{R}_{E}-1}$, $\gamma_{k,C}=2^{\bar{R}_{k,C}-1}$. Note that problem (\ref{p2}) is still non-convex due to the coupling in the SINR expressions and the rank-one constraints. To circumvent the obstacle, we convert the objective function into the following equivalent forms:
\begin{equation}
	\sum_{k=1}^{2}\log_2\delta_{k}-\sum_{k=1}^{2}\log_2\varphi_{k}+\log_2\psi-\log_2\phi,
\end{equation}
where $\delta_{k}$, $\varphi_{k}$, $\psi$, $\phi$ are slack optimization variables. Then, we add the new constraints with respect to $\delta_{k}$, $\varphi_{k}$, $\psi$, $\phi$, which are given by
\begin{equation}\label{delta_k}
	\delta_{k}\geq{\rm Tr}(\mathbf{U}_{r}\mathbf{F}_{k}\mathbf{W}_{k,E}\mathbf{F}_{k}^H)+\sigma_{k,C}^2,
\end{equation}
\begin{equation}\label{var_k}
	\varphi_{k}\leq{\rm Tr}(\mathbf{U}_{r}\mathbf{F}_{k}\mathbf{W}_{k,C}\mathbf{F}_{k}^H) +{\rm Tr}(\mathbf{U}_{r}\mathbf{F}_{k}\mathbf{W}_{k,E}\mathbf{F}_{k}^H)+\sigma_{k,C}^2,
\end{equation}
\begin{equation}\label{psi}
	\psi\geq{\rm Tr}(\mathbf{U}_{t}\mathbf{H}\mathbf{W}_{C}\mathbf{H}^H)+\sigma_{E}^2,
\end{equation}
and
\begin{equation}\label{phi}
	\phi\leq{\rm Tr}(\mathbf{U}_{t}\mathbf{H}\mathbf{W}_{C}\mathbf{H}^H)+{\rm Tr}(\mathbf{U}_{t}\mathbf{H}\mathbf{W}_{E}\mathbf{H}^H)+\sigma_{E}^2,
\end{equation}
respectively. It can be seen that (\ref{delta_k})–(\ref{phi}) have a similar structure consisting a product of matrices. In the following, we recast these equations in the form of difference of the convex (DC) functions. In particular, we tack as an example to demonstrate the logic of the proposed transformation. 
\textit{Theorem}
	For two arbitrary Hermitian matrices $\mathbf{A}\in \mathbb{H}^M$ and $\mathbf{B}\in \mathbb{H}^M$, we have
	\begin{equation}\label{TAB_1}
		\mathtt{Tr}(\mathbf{AB})=\frac{1}{2}\|\mathbf{A}+\mathbf{B}\|_F^2-\frac{1}{2}\|\mathbf{A}\|_F^2-\frac{1}{2}\|\mathbf{B}\|_F^2
	\end{equation}
	\begin{equation}\label{TAB_2}
		-\mathtt{Tr}(\mathbf{AB})=\frac{1}{2}\|\mathbf{A}-\mathbf{B}\|_F^2-\frac{1}{2}\|\mathbf{A}\|_F^2-\frac{1}{2}\|\mathbf{B}\|_F^2
	\end{equation}
\textit{Proof}
	The right side of (\ref{TAB_1}) can be expressed as
	\begin{align}\label{proof_TAB_1}
		&\frac{1}{2}\|\mathbf{A}+\mathbf{B}\|_F^2-\frac{1}{2}\|\mathbf{A}\|_F^2-\frac{1}{2}\|\mathbf{B}\|_F^2 \nonumber\\
		&=\frac{1}{2}\mathtt{Tr}\left( (\mathbf{A}+\mathbf{B})^H(\mathbf{A}+\mathbf{B}) \right)- \frac{1}{2}\mathtt{Tr}\left( \mathbf{A}^H\mathbf{A} \right)- \frac{1}{2}\mathtt{Tr}\left( \mathbf{B}^H\mathbf{B} \right) \nonumber\\
		&=\frac{1}{2}\mathtt{Tr}\left( \mathbf{A}^H\mathbf{B} \right)+\frac{1}{2}\mathtt{Tr}\left( \mathbf{B}^H\mathbf{A} \right)\overset{\mathrm{a}}{=}\mathtt{Tr}(\mathbf{A}\mathbf{B})
	\end{align}
	where the $\overset{\mathrm{a}}{=}$ holds when $\mathbf{A}$ and $\mathbf{B}$ are Hermitian matrices. Similar to (\ref{proof_TAB_1}), (\ref{TAB_2}) is also valid.
	
According to \textbf{Theorem 1}, the first and second non-convex terms
in (\ref{p2}c) can be transformed into the following difference of convex functions:
\begin{align}
	&\gamma_{k,C}{\rm Tr}(\mathbf{U}_{r}\mathbf{F}_{k}\mathbf{W}_{k,E}\mathbf{F}_{k}^H)=\frac{\gamma_{k,C}}{2}||\mathbf{U}_{r}+\mathbf{F}_{k}\mathbf{W}_{k,E}\mathbf{F}_{k}^H||_{F}^2 \nonumber\\
	&~~~-\frac{\gamma_{k,C}}{2}||\mathbf{U}_{r}||_{F}^2-\frac{\gamma_{k,C}}{2}||\mathbf{F}_{k}\mathbf{W}_{k,E}\mathbf{F}_{k}^H||_{F}^2
\end{align}
\begin{align}
	&-{\rm Tr}(\mathbf{U}_{r}\mathbf{F}_{k}\mathbf{W}_{k,C}\mathbf{F}_{k}^H)=\frac{1}{2}||\mathbf{U}_{r}-\mathbf{F}_{k}\mathbf{W}_{k,C}\mathbf{F}_{k}^H||_{F}^2\nonumber\\
	&~~~-\frac{1}{2}||\mathbf{U}_{r}||_{F}^2-\frac{1}{2}||\mathbf{F}_{k}\mathbf{W}_{k,C}\mathbf{F}_{k}^H||_{F}^2
\end{align}
where the terms $-||\mathbf{F}_{k}\mathbf{W}_{k,E}\mathbf{F}_{k}^H||_{F}^2$, $-||\mathbf{F}_{k}\mathbf{W}_{k,C}\mathbf{F}_{k}^H||_{F}^2$ and $-||\mathbf{U}_{r}||_{F}^2$ are concave with respect to $\mathbf{W}_{k,E}$, $\mathbf{W}_{k,C}$ and $\mathbf{U}_{r}$, respectively, while the term $||\mathbf{U}_{r}+\mathbf{F}_{k}\mathbf{W}_{k,E}\mathbf{F}_{k}^H||_{F}^2$ is convex with respect to $\mathbf{W}_{k,E}$ and $\mathbf{U}_{r}$, the term $||\mathbf{U}_{r}-\mathbf{F}_{k}\mathbf{W}_{k,C}\mathbf{F}_{k}^H||_{F}^2$ is convex with respect to $\mathbf{W}_{k,C}$ and $\mathbf{U}_{r}$. Then, we apply the SCA approach in each iteration to construct a global lower bound of the terms via their first-order Taylor approximations, respectively, which are given by
\begin{equation}
	||\mathbf{U}_{r}||_{F}^2\geq-||\mathbf{U}_{r}^{(t)}||_{F}^2+2{\rm Tr}\left((\mathbf{U}_{r}^{(t)})^H\mathbf{U}_{r}\right)
\end{equation}
\begin{align}
	&||\mathbf{F}_{k}\mathbf{W}_{k,E}\mathbf{F}_{k}^H||_{F}^2\geq-||\mathbf{F}_{k}\mathbf{W}_{k,E}^{(t)}\mathbf{F}_{k}^H||_{F}^2 \nonumber\\
	&~~~+2{\rm Tr}\left((\mathbf{F}_{k}^H\mathbf{F}_{k}\mathbf{W}_{k,E}^{(t)}\mathbf{F}_{k}^H\mathbf{F}_{k})^H\mathbf{W}_{k,E}\right)
\end{align}
and
\begin{align}
	&||\mathbf{F}_{k}\mathbf{W}_{k,C}\mathbf{F}_{k}^H||_{F}^2\geq-||\mathbf{F}_{k}\mathbf{W}_{k,C}^{(t)}\mathbf{F}_{k}^H||_{F}^2\nonumber\\
	&~~~+2{\rm Tr}\left((\mathbf{F}_{k}^H\mathbf{F}_{k}\mathbf{W}_{k,C}^{(t)}\mathbf{F}_{k}^H\mathbf{F}_{k})^H\mathbf{W}_{k,C}\right)
\end{align}
respectively, where $\mathbf{U}_{r}^{(t)}$, $\mathbf{W}_{k,E}^{(t)}$ and $\mathbf{W}_{k,C}^{(t)}$ are feasible points in the $t$-th iteration of the SCA. As a result, the non-convex constraint is replace by
\begin{align}\label{p1_c1}
	&\frac{\gamma_{k,C}}{2}||\mathbf{U}_{r}+\mathbf{F}_{k}\mathbf{W}_{k,E}\mathbf{F}_{k}^H||_{F}^2-(\gamma_{k,C}+1){\rm Tr}\left((\mathbf{U}_{r}^{(t)})\mathbf{U}_{r}\right)\notag\\
	&+\frac{\gamma_{k,C}+1}{2}||\mathbf{U}_{r}^{(t)}||_{F}^2-\gamma_{k,C}{\rm Tr}\left((\mathbf{F}_{k}^H\mathbf{F}_{k}\mathbf{W}_{k,E}^{(t)}\mathbf{F}_{k}^H\mathbf{F}_{k})^H\mathbf{W}_{k,E}\right)\notag\\
	&+\frac{\gamma_{k,C}}{2}||\mathbf{F}_{k}\mathbf{W}_{k,E}^{(t)}\mathbf{F}_{k}^H||_{F}^2+\frac{1}{2}||\mathbf{U}_{r}-\mathbf{F}_{k}\mathbf{W}_{k,C}\mathbf{F}_{k}^H||_{F}^2\notag\\
	&-{\rm Tr}\left((\mathbf{F}_{k}^H\mathbf{F}_{k}\mathbf{W}_{k,C}^{(t)}\mathbf{F}_{k}^H\mathbf{F}_{k})^H\mathbf{W}_{k,C}\right)+\frac{1}{2}||\mathbf{F}_{k}\mathbf{W}_{k,C}^{(t)}\mathbf{F}_{k}^H||_{F}^2\notag\\
	&+\gamma_{k,C}\sigma_{k,C}^2\leq0
\end{align}
Then, the constraint (\ref{p2}d) can be handled via a similar manner as. Therefore, the non-convex constraint is replaced by
\begin{align}\label{p1_c2}
	&\frac{\gamma_{E}}{2}||\mathbf{U}_{t}+\mathbf{H}\mathbf{W}_{C}\mathbf{H}^H||_{F}^2-(\gamma_{E}+1){\rm Tr}\left((\mathbf{U}_{t}^{(t)})\mathbf{U}_{t}\right)\notag\\
	&+\frac{\gamma_{E}+1}{2}||\mathbf{U}_{t}^{(t)}||_{F}^2-\gamma_{E}{\rm Tr}\left((\mathbf{H}^H\mathbf{H}\mathbf{W}_{C}^{(t)}\mathbf{H}^H\mathbf{H})^H\mathbf{W}_{C}\right)\notag\\
	&+\frac{\gamma_{E}}{2}||\mathbf{H}\mathbf{W}_{C}^{(t)}\mathbf{H}^H||_{F}^2+\frac{1}{2}||\mathbf{U}_{t}-\mathbf{H}\mathbf{W}_{E}\mathbf{H}^H||_{F}^2\notag\\
	&-{\rm Tr}\left((\mathbf{H}^H\mathbf{H}\mathbf{W}_{E}^{(t)}\mathbf{H}^H\mathbf{H})^H\mathbf{W}_{E}\right)+\frac{1}{2}||\mathbf{H}\mathbf{W}_{E}^{(t)}\mathbf{H}^H||_{F}^2\notag\\
	&+\gamma_{E}\sigma_{E}^2\leq0
\end{align}
Similarly, the constraints (\ref{delta_k})–(\ref{phi}) are replaced by
\begin{align}\label{p1_c3}
	&\frac{1}{2}||\mathbf{U}_{r}+\mathbf{F}_{k}\mathbf{W}_{k,E}\mathbf{F}_{k}^H||_{F}^2-{\rm Tr}\left((\mathbf{U}_{r}^{(t)})\mathbf{U}_{r}\right)\notag\\
	&+\frac{1}{2}||\mathbf{U}_{r}^{(t)}||_{F}^2-{\rm Tr}\left((\mathbf{F}_{k}^H\mathbf{F}_{k}\mathbf{W}_{k,E}^{(t)}\mathbf{F}_{k}^H\mathbf{F}_{k})^H\mathbf{W}_{k,E}\right)\notag\\
	&+\frac{1}{2}||\mathbf{F}_{k}\mathbf{W}_{k,E}^{(t)}\mathbf{F}_{k}^H||_{F}^2+\sigma_{k,C}^2-\delta_{k}\leq0
\end{align}
\begin{align}\label{p1_c4}
	&\frac{1}{2}||\mathbf{U}_{r}-\mathbf{F}_{k}\mathbf{W}_{k,C}\mathbf{F}_{k}^H||_{F}^2-2{\rm Tr}\left((\mathbf{U}_{r}^{(t)})\mathbf{U}_{r}\right)\notag\\
	&+||\mathbf{U}_{r}^{(t)}||_{F}^2-{\rm Tr}\left((\mathbf{F}_{k}^H\mathbf{F}_{k}\mathbf{W}_{k,C}^{(t)}\mathbf{F}_{k}^H\mathbf{F}_{k})^H\mathbf{W}_{k,C}\right)\notag\\
	&+\frac{1}{2}||\mathbf{F}_{k}\mathbf{W}_{k,C}^{(t)}\mathbf{F}_{k}^H||_{F}^2+\frac{1}{2}||\mathbf{U}_{r}-\mathbf{F}_{k}\mathbf{W}_{k,E}\mathbf{F}_{k}^H||_{F}^2\notag\\
	&-{\rm Tr}\left((\mathbf{F}_{k}^H\mathbf{F}_{k}\mathbf{W}_{k,E}^{(t)}\mathbf{F}_{k}^H\mathbf{F}_{k})^H\mathbf{W}_{k,E}\right)+\frac{1}{2}||\mathbf{F}_{k}\mathbf{W}_{k,E}^{(t)}\mathbf{F}_{k}^H||_{F}^2\notag\\
	&+\sigma_{k,C}^2+\varphi_{k}\leq0
\end{align}
\begin{align}\label{p1_c5}
	&\frac{1}{2}||\mathbf{U}_{t}+\mathbf{H}\mathbf{W}_{C}\mathbf{H}^H||_{F}^2-{\rm Tr}\left((\mathbf{U}_{t}^{(t)})\mathbf{U}_{t}\right)\notag\\
	&+\frac{1}{2}||\mathbf{U}_{t}^{(t)}||_{F}^2-{\rm Tr}\left((\mathbf{H}^H\mathbf{H}\mathbf{W}_{C}^{(t)}\mathbf{H}^H\mathbf{H})^H\mathbf{W}_{C}\right)\notag\\
	&+\frac{1}{2}||\mathbf{H}\mathbf{W}_{C}^{(t)}\mathbf{H}^H||_{F}^2+\sigma_{E}^2-\psi\leq0
\end{align}
\begin{align}\label{p1_c6}
	&\frac{1}{2}||\mathbf{U}_{t}-\mathbf{H}\mathbf{W}_{C}\mathbf{H}^H||_{F}^2-2{\rm Tr}\left((\mathbf{U}_{t}^{(t)})\mathbf{U}_{t}\right)\notag\\
	&+||\mathbf{U}_{t}^{(t)}||_{F}^2-{\rm Tr}\left((\mathbf{H}^H\mathbf{H}\mathbf{W}_{C}^{(t)}\mathbf{H}^H\mathbf{H})^H\mathbf{W}_{C}\right)\notag\\
	&+\frac{1}{2}||\mathbf{H}\mathbf{W}_{C}^{(t)}\mathbf{H}^H||_{F}^2+\frac{1}{2}||\mathbf{U}_{t}-\mathbf{H}\mathbf{W}_{E}\mathbf{H}^H||_{F}^2\notag\\
	&-{\rm Tr}\left((\mathbf{H}^H\mathbf{H}\mathbf{W}_{E}^{(t)}\mathbf{H}^H\mathbf{H})^H\mathbf{W}_{E}\right)+\frac{1}{2}||\mathbf{H}\mathbf{W}_{E}^{(t)}\mathbf{H}^H||_{F}^2\notag\\
	&+\sigma_{E}^2+\phi\leq0
\end{align}
respectively.
Next, we address the non-convex rank-one constraint (\ref{p2}j). The constraint (\ref{p2}j) can be transform into the following equivalent forms:
\begin{equation}\label{U_l}
	||\mathbf{U}_{l}||_{*}-||\mathbf{U}_{l}||_{2}\leq0,\enspace\forall l,
\end{equation}
where $||\mathbf{U}_{l}||_{*}=\sum\nolimits_{i}\sigma_{i}(\mathbf{U}_{l})$, $||\mathbf{U}_{l}||_{2}=\max\limits_{i}\sigma_{i}(\mathbf{U}_{l})$ denote the nuclear norm and the spectral norm, respectively, and $\sigma_{i}(\mathbf{U}_{l})$ is the $i$-th largest singular value of $\mathbf{U}_{l}$. Note that we always have $||\mathbf{U}_{l}||_{*}-||\mathbf{U}_{l}||_{2}\geq0$, where equality holds if and only if $\mathbf{U}_{l}$ is a rank-one matrix. Hence, equality constraint (\ref{U_l}) is only met for rank-one matrices ${\mathbf{U}_{l}}$. Note that the constraint (\ref{U_l}) is still non-convex. Next, we adopt the penalty method to solve this issue. By augmenting the constraint (\ref{U_l}) into the objective function, we obtain the following problem:
\begin{subequations}\label{p3}
	\begin{align}
		\min\limits_{\mathbf{W}_{k,C},\mathbf{W}_{k,E},\mathbf{W}_{C},\atop\mathbf{W}_{E},\mathbf{U}_{l},\delta_{k},\varphi_{k},\psi,\phi}\enspace &f_{\text{CU}}-g_{\text{CU}}+f_{\text{CE}}-g_{\text{CE}} \nonumber\\
		&+\chi\sum\nolimits_{l}(||\mathbf{U}_{l}||_{*}-||\mathbf{U}_{l}||_{2})\\
		\text {s.t.}\enspace&\text{(\ref{p1_c1}-\ref{p1_c6})},\\
		&\text{(\ref{p2}e-\ref{p2}i)}
	\end{align}
\end{subequations}
where $f_{\text{CU}}=\sum_{k=1}^{2}\log_{2}\delta_{k}$, $g_{\text{CU}}=\sum_{k=1}^{2}\log_{2}\varphi_{k}$, $f_{\text{CE}}=\log_{2}\psi$ and $g_{\text{CE}}=\log_{2}\phi$. The constraint (\ref{U_l}) is relaxed to a penalty term added to the objective function, and $\chi > 0$ is the penalty factor which penalizes the objective function for any $\mathbf{U}_{l}$ whose rank is larger than one. According to , the problems (\ref{p2}) and (\ref{p3}) are equivalent when $\chi \rightarrow +\infty$. Then, we initialize $\chi$ with a small value to find a good starting point, and we gradually increase $\chi$ to a sufficiently large value in the course of several iterations to finally obtain feasible rank-one matrices. Note that, for any given penalty factor $\chi > 0$, problem (\ref{p3}) is still non-convex due to the non-convexity of the objective function and the constraints.

Based on the proposed transformations, the objective function is also in the form of DC. To handle the non-convex objective function, for any feasible points $\delta_{k}^{(t)}$ and $\psi^{(t)}$ in the $t$-th iteration of SCA, using first-order Taylor expansion, the global lower bounds of $f_{\text{CU}}$ and $f_{\text{CE}}$ can be obtained as follows:
\begin{equation}
	f_{\text{CU}}\geq\sum_{k=1}^{2}(\log_{2}\delta_{k}^{(t)}+\frac{\delta_{k}-\delta_{k}^{(t)}}{\ln{2}\delta_{k}^{(t)}})\triangleq\bar{f}_{\text{CU}}(\delta_{k},\delta_{k}^{(t)}),
\end{equation}
\begin{equation}
	f_{\text{CE}}\geq\log_{2}\psi^{(t)}+\frac{\psi-\psi^{(t)}}{\ln{2}\psi^{(t)}}\triangleq\bar{f}_{\text{CE}}(\psi,\psi^{(t)}),
\end{equation}
Similarly, for any feasible points $\mathbf{U}_{l}^{(t)}$ in the $t$-th iteration, we obtain a convex upper bound for the penalty term, which is given by
\begin{equation}
	||\mathbf{U}_{l}||_{*}-||\mathbf{U}_{l}||_{2}\leq||\mathbf{U}_{l}||_{*}-\bar{\mathbf{U}}_{l}^{(t)}
\end{equation}
where
\begin{equation}
	\bar{\mathbf{U}}_{l}^{(t)}\triangleq||\bar{\mathbf{U}}_{l}^{(t)}||_{2}+{\rm Tr}\left(\bm{\lambda}(\mathbf{U}_{l}^{(t)})(\bm{\lambda}(\mathbf{U}_{l}^{(t)}))^H(\mathbf{U}_{l}-\mathbf{U}_{l}^{(t)}\right)
\end{equation}
$\bm{\lambda}(\mathbf{U}_{l}^{(t)})$ denotes the eigenvector corresponding to the largest eigenvalue of $\mathbf{U}_{l}^{(t)}$. Then, problem (\ref{p3}) is replaced by
\begin{subequations}\label{p4}
	\begin{align}
		\min\limits_{\mathbf{W}_{k,C},\mathbf{W}_{k,E},\mathbf{W}_{C},\atop\mathbf{W}_{E},\mathbf{U}_{l},\delta_{k},\varphi_{k},\psi,\phi}\enspace &\bar{f}_{\text{CU}}-g_{\text{CU}}+\bar{f}_{\text{CE}}-g_{\text{CE}}
		\nonumber\\
		&+\chi\sum\nolimits_{l}(||\mathbf{U}_{l}||_{*}-\bar{\mathbf{U}}_{l}^{(t)})\\
		\text {s.t.}\enspace&\text{(\ref{p1_c1}-\ref{p1_c6})},\\
		&\text{(\ref{p2}e-\ref{p2}i)}
	\end{align}
\end{subequations}
Now, the remaining non-convexity of (\ref{p4}) is rank-one constraints (\ref{p2}i). We employ SDR and solve the relaxed problem by ignoring (\ref{p2}i). The relaxed problem (\ref{p4}), which is a standard convex semidefinite program (SDP), can be efficiently solved via standard convex problem solvers such as CVX \cite{cvx}. Then, we propose a penalty-based iterative algorithm, which comprises two loops. In the outer loop, the penalty factor is gradually increased as follows: $\chi=\omega\chi$, where $\omega>1$. The algorithm terminates when the penalty term satisfies the following criterion:
\begin{equation}\label{conv_cons1}
	\max\left\{ ||\mathbf{U}_l||_* - ||\mathbf{U}_l||_2,\forall l \in \{t,r\}\right\} \le \varepsilon _1
\end{equation}
where $\varepsilon _1$ denotes a predefined maximum violation of constraint (\ref{U_l}). Therefore, (\ref{U_l}) will be finally satisfied
with accuracy $\varepsilon _1$ as $\chi$ increases. In the inner loop, variables are jointly optimized by iteratively solving the relaxed problem (\ref{p4}) for the given penalty factor. The objective function value of the (\ref{p4}) is non-increasing
in each iteration of the inner loop. Therefore, as $\chi$ approaches infinity, the developed penalty-based iterative algorithm is guaranteed to converge to a stationary point of the original problem (\ref{p2}). 

%

\subsection{Joint Design of Beamforming Vector and STAR-RIS Coefficient Matrix for MS Protocol}
After solving the optimization problem in ES mode, this subsection proceeds to study the sum rate and maximization problem in MS protocol, where the magnitude coefficients of STAR-RIS can only be chosen between 0 and 1. Therefore, the optimization problem on the STAR-RIS coefficients becomes a 0-1 integer programming problem, which can be given by
\begin{subequations}\label{p21}
	\begin{align}
		&\max\limits_{\mathbf{w}_{k,C},\mathbf{w}_{k,E},\mathbf{u}_{l}}\enspace\sum_{k=1}^{2}R_{k,C}+R_{E}\\
		\text {s.t.}\enspace&\enspace||\mathbf{w}_{k,C}||^2+||\mathbf{w}_{k,E}||^2\leq P,\enspace\forall k\\
		&\enspace R_{k,C}\geq \bar{R}_{k,C},\enspace\forall k\\
		&\enspace R_{E}\geq \bar{R}_{E}\\
		&\enspace[\mathbf{u}_{l}]_{n}=\sqrt{\beta_{n}^{l}}e^{j\theta_{n}^{l}},\theta_{n}^{l}\in[0,2\pi),\enspace\forall l,n\\
		&\enspace[\mathbf{u}_{l}]_{N+1}=1\\
		&\enspace\beta_{n}^{l}\in\{0,1\},\enspace\forall l,n
	\end{align}
\end{subequations}
Compared with the ES protocol, the optimization problem involves an additional non-convex binary constraint, $\beta_{n}^{l}\in\{0,1\}$, in the MS protocol. Since other nonconvex items can be handled in a similar way as in the previous subsection, it is only necessary to focus on how to solve this new binary constraint. This binary constraint could be transferred into a equation constraint as follow
\begin{equation}\label{equ_cons}
	\beta_{n}^{l}-(\beta_{n}^{l})^2=0,\forall l,n
\end{equation}
Because the amplitude coefficients of transmission and reflection of STAR-RIS take values between 0 and 1, $\beta_{n}^{l}-(\beta_{n}^{l})^2\geq0$ is always satisfied and the equality holds if and only if $\beta_{n}^{l}=0$ or $\beta_{n}^{l}=1$. Thus, (\ref{equ_cons}) is satisfied when $\beta_{n}^{l}=0$ or $\beta_{n}^{l}=1$.

In the following, we will investigate how to extend the penalty function based iterative algorithm proposed in the ES protocol to the optimization problem in the MS protocol. Based on the method for ES protocol, (\ref{equ_cons}) can be added to the objective function as an additional penalty term. Thus, (\ref{p21}) can be transformed into
\begin{subequations}\label{p22}
	\begin{align}
		\min\limits_{\mathbf{W}_{k,C},\mathbf{W}_{k,E},\mathbf{W}_{C},\atop\mathbf{W}_{E},\mathbf{U}_{l},\delta_{k},\varphi_{k},\psi,\phi}\enspace &\bar{f}_{\text{CU}}-g_{\text{CU}}+\bar{f}_{\text{CE}}-g_{\text{CE}}
		\nonumber\\
		&+\chi\sum\nolimits_{l}(||\mathbf{U}_{l}||_{*}-\bar{\mathbf{U}}_{l}^{(t)})\nonumber\\
		&+\eta\sum_{n=1}^{N}\sum\nolimits_{l}\left(\beta_{n}^{l}-(\beta_{n}^{l})^2\right) \\
		\text {s.t.}\enspace&\text{(\ref{p1_c1}-\ref{p1_c6})},\\
		&\text{(\ref{p2}e-\ref{p2}i)}
	\end{align}
\end{subequations}
where $\eta>0$ denotes a new penalty factor. When $\chi,\eta\rightarrow + \infty$, the solution of (\ref{p22}) is able to satisfy (\ref{U_l}) and (\ref{equ_cons}). Then, the SCA can still be adopted to solve this nonconvex optimization problem. For a given point in the $t$th iteration of the SCA, $\{\beta_l^{(t)}\}$, an upper bound for the new penalty term could be achieved via using the first-order Taylor approximation
\begin{align}\label{beta_up}
	\beta_{n}^{l}-(\beta_{n}^{l})^2&\geq\beta_{n}^{l}-(\beta_{n}^{l(t)})^2-2\beta_{n}^{l(t)}\left(\beta_{n}^{l}-\beta_{n}^{l(t)}\right) \nonumber\\
	&=\left(1-2\beta_{n}^{l(t)}\right)\beta_{n}^{l}+\left(\beta_{n}^{l(t)}\right)^2 \nonumber\\
	&\triangleq\Omega\left(\beta_{n}^{l},\beta_{n}^{l(t)}\right), \forall l,n
\end{align}
Substituting (\ref{beta_up}) into (\ref{p22}a), we have
\begin{subequations}\label{p23}
	\begin{align}
		\min\limits_{\mathbf{W}_{k,C},\mathbf{W}_{k,E},\mathbf{W}_{C},\atop\mathbf{W}_{E},\mathbf{U}_{l},\delta_{k},\varphi_{k},\psi,\phi}\enspace &\bar{f}_{\text{CU}}-g_{\text{CU}}+\bar{f}_{\text{CE}}-g_{\text{CE}}
		\nonumber\\
		&+\chi\sum\nolimits_{l}(||\mathbf{U}_{l}||_{*}-\bar{\mathbf{U}}_{l}^{(t)})\nonumber\\
		&+\eta\sum_{n=1}^{N}\sum\nolimits_{l}\Omega\left(\beta_{n}^{l},\beta_{n}^{l(t)}\right) \\
		\text {s.t.}\enspace&\text{(\ref{p1_c1}-\ref{p1_c6})},\\
		&\text{(\ref{p2}e-\ref{p2}i)}
	\end{align}
\end{subequations}
Then, SDR is used again to remove the nonconvex rank-one constraint, and the relaxed problem is also convex and can be solved directly with CVX \cite{cvx}. Similar to the ES protocol, the two-level cyclic penalty function based iterative algorithm can still be used for the optimization problem in the MS protocol. Since there are two penalty factors in (\ref{p23}), the algorithm terminates when the penalty terms satisfy the following criterion:
\begin{equation}\label{conv_cons2}
	\max\left\{
	\begin{array}{c}
		||\mathbf{U}_l||_* - ||\mathbf{U}_l||_2 \\
		\beta_{n}^{l}-(\beta_{n}^{l})^2
	\end{array}
	\right\}\geq\varepsilon_2,\forall l \in \{t,r\}
\end{equation}
where $\varepsilon_2$ denotes a predefined maximum violation of constraint (\ref{U_l}) and (\ref{beta_up}). 

%

\section{Analysis of Computational Complexity}\label{sec_com}
In this section, we analyze the computational complexity of the proposed penalty function based iterative algorithm. The main complexity lies in solving the SDR after relaxation in the inner loop, and since the relaxation problem is a standard SDP problem, the computational complexity of solving this problem is given by
\begin{equation}
	\mathcal{C}_s = \mathcal{O}\left(KM^{6.5}+2N^{6.5}\right)
\end{equation}
Therefore, the total computational complexity of the proposed penalty function based iterative algorithm can be expressed by
relaxation problem is a standard SDP problem, the computational complexity of solving this problem is given by
\begin{equation}
	\mathcal{C}_p = \mathcal{O}\left(I_{out}I_{in}\left(KM^{6.5}+2N^{6.5}\right)\right)
\end{equation}
where $I_{out}$ and $I_{in}$ denote the number of iterations required for convergence of inner and outer loops, respectively. It can be seen that the computational complexity of this algorithm is a polynomial over $N$. Thus, even if the number of elements of STAR-RIS is too large, the algorithm still has a relative low computational complexity. Thus, it is easy to be implemented in practical applications.

\section{Simulation Results and Analysis}\label{sec_simu}
In this section, the performance of the STAR-RIS-UAV aided CoMP communication system is simulated and verified. Consider a three-dimensional coordinate system, 2 BSs are considered and located at (0m, 0m, 10m) and (140m, 0m, 10m), respectively, with the radius of each cell set to 80 m. 
STAR-RIS-UAV is located in the overlapping area of the two cells, set at (70m, 20m, 20m). 
CCU1 and CCU2 are located in the reflection half of the STAR-RIS and are randomly distributed in the two discs centered on the adjacent BS with the inner and outer diameters of 20m and 40m, respectively.
The CEU is randomly distributed in the overlapping area of the two cells and is located in the transmission half of the area 20m away from the STAR-RIS. The path loss is calculated as $PL=PL_0(d/d_0)^{-\alpha}$, where $PL_0=10^{-3}$ is the channel gain when $d_0=1$m, $d$ denotes the path distance, and $\alpha=2.2$ is the path loss index. In the simulation, all channel models use the Rice fading channel, and the channel  can be expressed as
\begin{equation}
	\bm{G}=\sqrt{\frac{\rho}{\rho+1}}\bm{G}^{LOS}+\sqrt{\frac{1}{\rho+1}}\bm{G}^{NLOS}
\end{equation}
where, $\rho=5$ dB is the Rice coefficient. $\bm{G}^{LOS}$ denotes the LOS component of the channel, obtained by multiplying the transmit and receive direction vectors. And, $\bm{G}^{NLOS}$ is the NLOS component, modeled by Rayleigh Fading. Given the maximum transmitting power of the two BSs is the same, and the minimum required rate $\bar{R}_{k,C}$ and $\bar{R}_{E}$ are equal, $\bar{R}_{k,C}=\bar{R}_{E}$. Noise variance is set as $\sigma_{k,C}^2=\sigma_{E}^2=-90$dBm. In addition, $\varepsilon_1=\varepsilon_2=10^{-7}$ and initial values of penalty factors, $\chi$ and $\eta$, are set as $10^{-7}$.

In order to better investigate the performance of STAR-RIS in the corresponding ES and MS protocols, three comparison schemes are proposed as follow
\begin{enumerate}
	\item Conventional RIS: Instead of using STAR-RIS, full coverage is achieved by a reflective-only  RIS and a transmissive-only RIS. The two conventional RISs are adjacent to each other and deployed at the same locations as the STAR-RIS. For a fair comparison, it is assumed that each conventional reflection/transmission RIS has $N/2$ elements, where the number has been set to even for simplicity. This comparison scheme can be considered as a special case of the STAR-RIS in MS mode, where $N/2$ elements of the STAR-RIS perform the reflecting function and the remaining elements perform the transmitting function.
	\item Uniform energy splitting (UES): It is assumed that the TARCs of all components of STAR-RIS in ES mode are equal, $\beta_{n}^t = \bar{\beta}_{n}^t$, $\beta_{n}^r = \bar{\beta}_{n}^r$, $\forall n$, where $0\leq\bar{\beta}_{n}^t,\bar{\beta}_{n}^r\leq 1$, $\bar{\beta}_{n}^t+\bar{\beta}_{n}^r=1$. UES can be seen as a special case of STAR-RIS in ES protocol, with a group/face amplitude design.
	\item No STAR-RIS: Without the assistance of STAR-RIS, the system becomes a traditional collaborative multi-point transmission communication system
\end{enumerate}

\begin{figure}
	\includegraphics[width=0.48\textwidth]{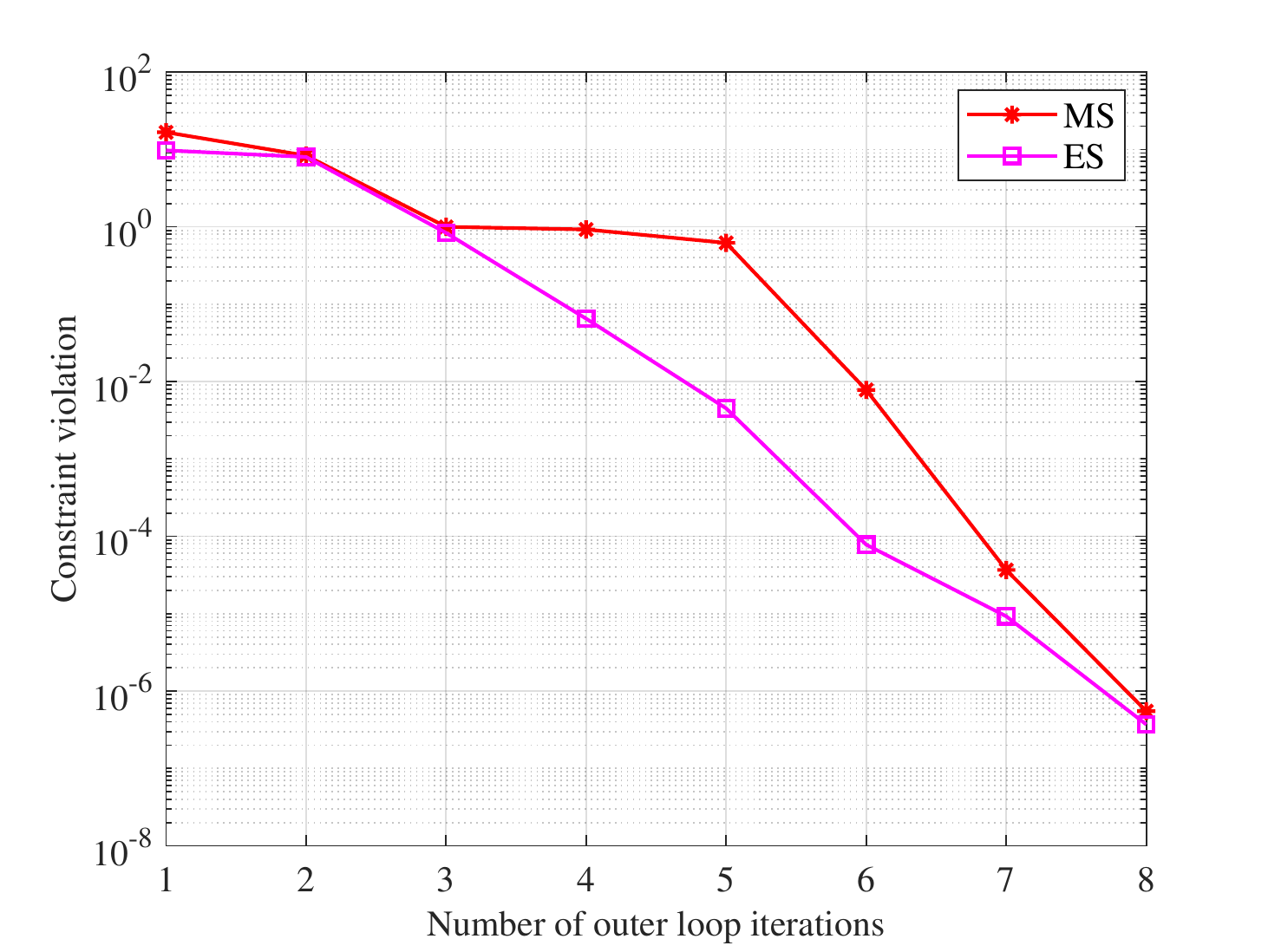}
	\caption{Constraint violation over the number of outer loop iterations .\label{fig_constraint}}
\end{figure}   

The constraint violation over the number of outer loop iterations for the penalty function-based iterative algorithms for ES and MS protocols are shown in Figure \ref{fig_constraint}. From Figure \ref{fig_constraint}, it can be seen that the constraint conflicts for both ES and MS decrease rapidly as the number of outer loop iterations increases and eventually satisfy the predetermined accuracy (i.e. $\varepsilon_1=\varepsilon_2=10^{-7}$) after 8 iterations. This implies that the feasible rank-one TARC matrices $\{\bm{U}_l\}$, as well as the binary TARCs $\{\beta_n^l\}$, have been optimized by the penalty function-based iterative algorithm.

\begin{figure}[!tb]
	\includegraphics[width=0.48\textwidth]{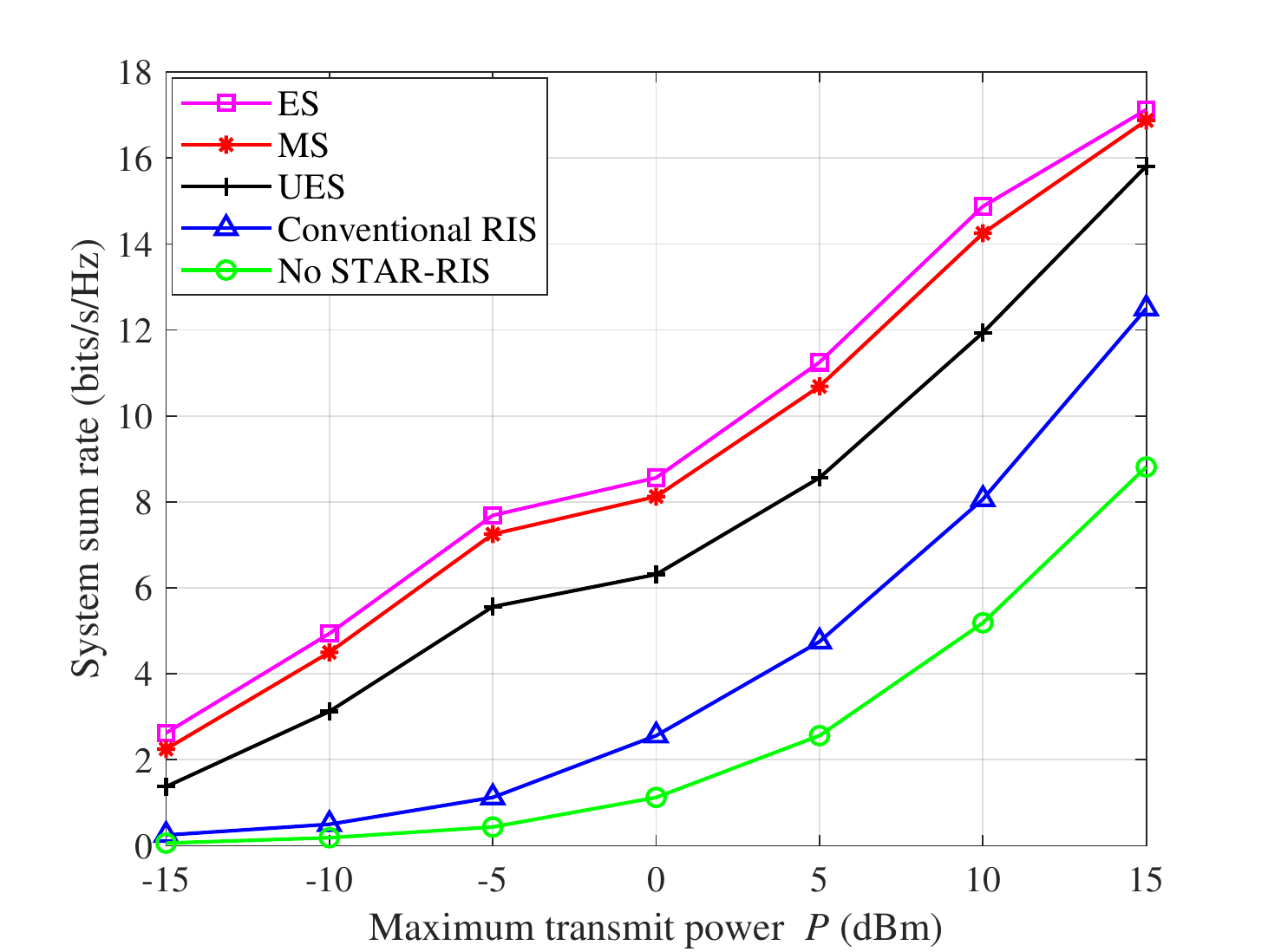}
	\caption{System sum rate versus the maximum transmit power.\label{fig_2p}}
\end{figure}   
\begin{figure}[!tb]
	\includegraphics[width=0.48\textwidth]{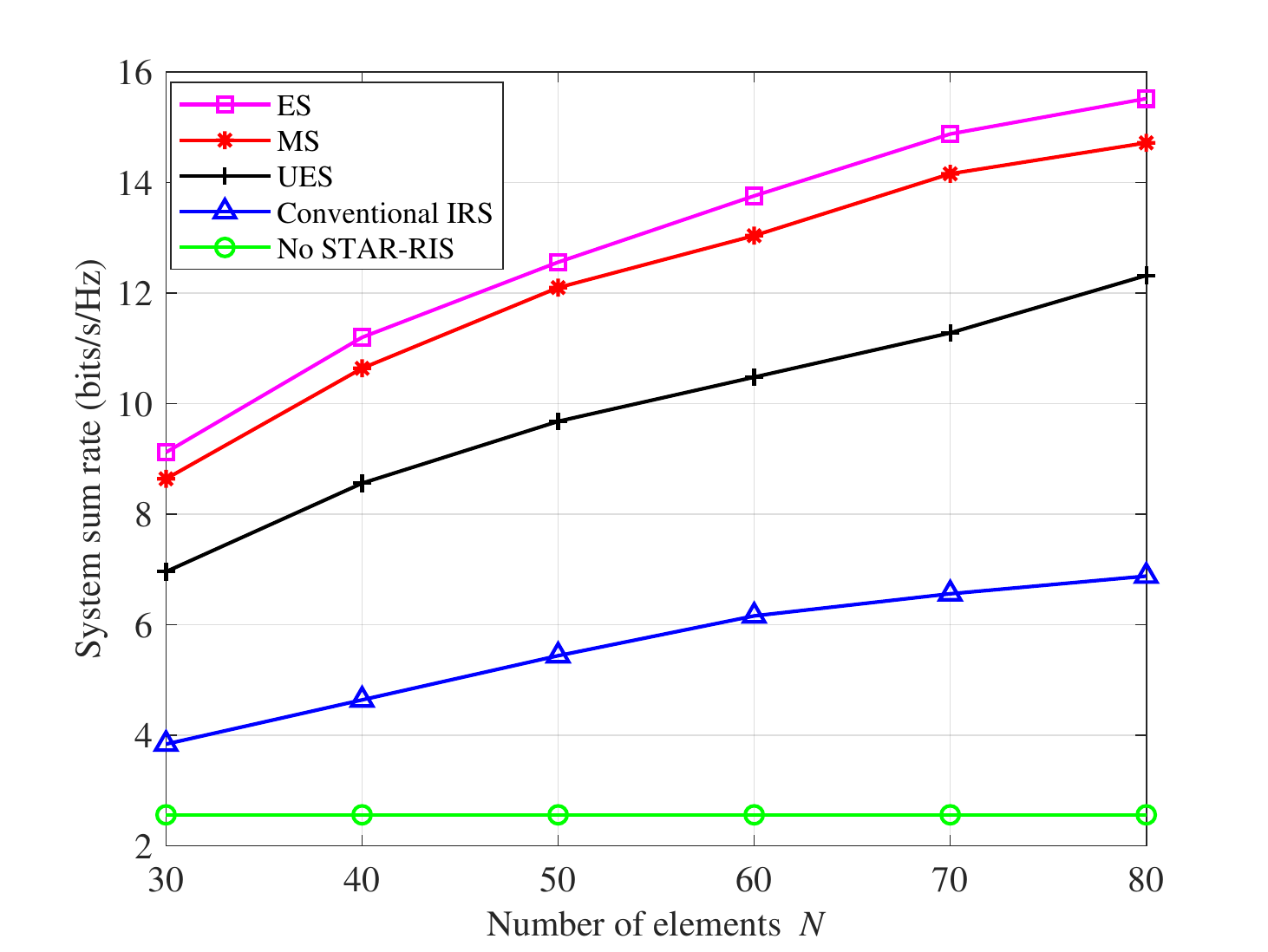}
	\caption{System sum rate versus the number of elements.\label{fig_2n}}
\end{figure}  
Figure \ref{fig_2p} shows a graph of the variation of the system sum rate versus the maximum transmit power. It can be seen that the system sum rate of all schemes increases monotonically with the increase of In addition, the ES mode obtains a higher system sum rate than the MS and UES schemes because the ES mode has a freer choice of amplitude and phase shift, while the MS and UES schemes have certain restrictions on the TARCs. the amplitude coefficients in MS can only be chosen from 0 and 1, while UES requires equal amplitude coefficients for all components. Therefore, the ES mode allows better adjustment of the amplitude and phase shift coefficients of each element, making full use of the degrees of freedom obtained by each element boost. from the mathematical and optimization point of view, both the MS mode and the UES scheme are a special case of the ES mode. In addition, for conventional RIS schemes, since only a fixed number of RIS elements perform a single reflection or transmission function, while STAR-RIS can flexibly control the number of elements involved in transmission and reflection, STAR-RIS can provide a greater degree of freedom than conventional RIS to enhance the signal strength, and therefore, conventional RIS schemes suffer some performance loss, especially when compared with ES and MS mode. In addition, it can be seen that the performance of the STAR-free RIS is the worst among these schemes. Through the effective combination of STAR-RIS and collaborative multipoint, the passive beamforming of STAR-RIS can be reasonably designed to enhance the combined channel gain at CEU and eliminate the adverse effects of inter-cell interference, thus further improving the system performance.

Figure \ref{fig_2n} depicts the variation of the system sum rate versus the number of STAR-RIS elements. From this figure, it can be observed that only the system sum rate of the no-STAR-RIS scheme remains constant, while the system sum rate of all the remaining schemes increases with the increase of the number of STAR-RIS elements. This is due to the fact that when the number of STAR-RIS elements increases, more signal energy can reach the STAR-RIS, so that the STAR-RIS elements can use more elements to control the energy and phase of the incident signal and optimize the transmission and reflection coefficients of each element rationally, thus increasing the system sum rate. Installing more elements also provides more freedom in resource allocation, which facilitates higher beamforming gain and thus higher system throughput. More importantly, since the reflective elements are passive, low power consumption and low cost, STAR-RIS with hundreds or even thousands of elements is expected to be used. it can also be seen from the figure that the performance of ES mode and MS is still better than other comparison schemes, once again proving the superiority of STAR-RIS.

\begin{figure}
	\includegraphics[width=0.48\textwidth]{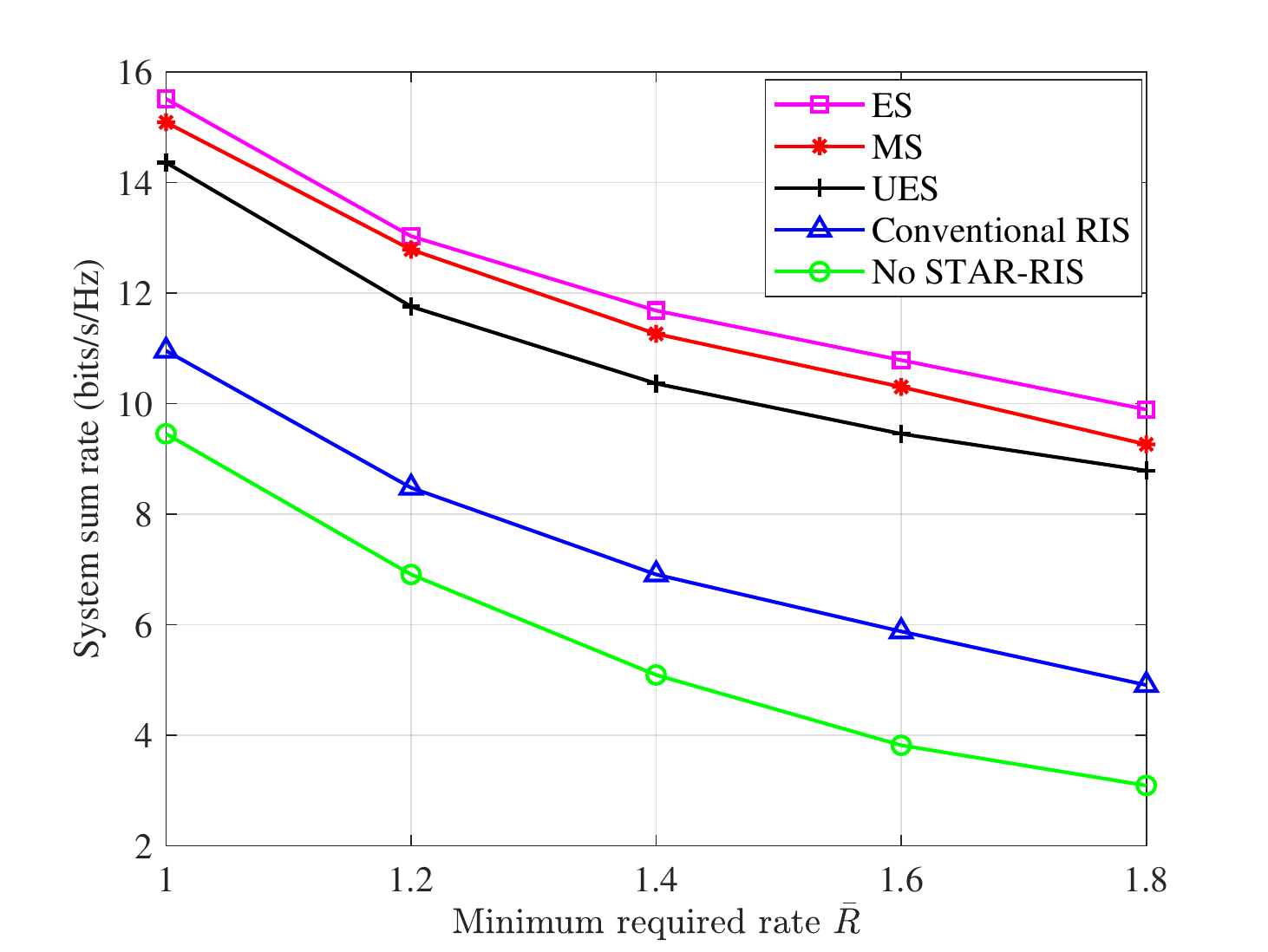}
	\caption{System sum rate over the minimum required rate\label{fig_r}}
\end{figure} 

Figure \ref{fig_r} plots the system sum rate over the variation with the minimum required rate. As observed in Figure \ref{fig_r}, the system sum rate and decreases monotonically as the minimum required rate of the user increases. In particular, when the user's channel conditions are poor, the increase in the minimum required rate causes the system to be forced to allocate more resources to it, reducing the flexibility of the communication system resource allocation. In addition, compared with the no-STAR-RIS scheme, the system rate and the decreasing rate with increasing minimum required rate are relatively slow in the scheme with STAR-RIS as well as the conventional RIS assistance. This is because the deployment of RIS can mitigate the effect of interference in the system, and without the assistance of STAR-RIS, the reflected signal cannot accurately reach the subscriber side to promote effective beamforming. The system is more sensitive to minimum rate requirements, and the additional degrees of freedom brought by collaborative multipoint techniques cannot be fully used to solve the inter-cell interference problem. Similar to the results in Figure \ref{fig_2n} and Figure \ref{fig_r}, the ES and MS modes with STAR-RIS still achieve the best performance, highlighting the importance of STAR-RIS as an aid to collaborative multipoint transmission systems.

\begin{figure}
	\includegraphics[width=0.48\textwidth]{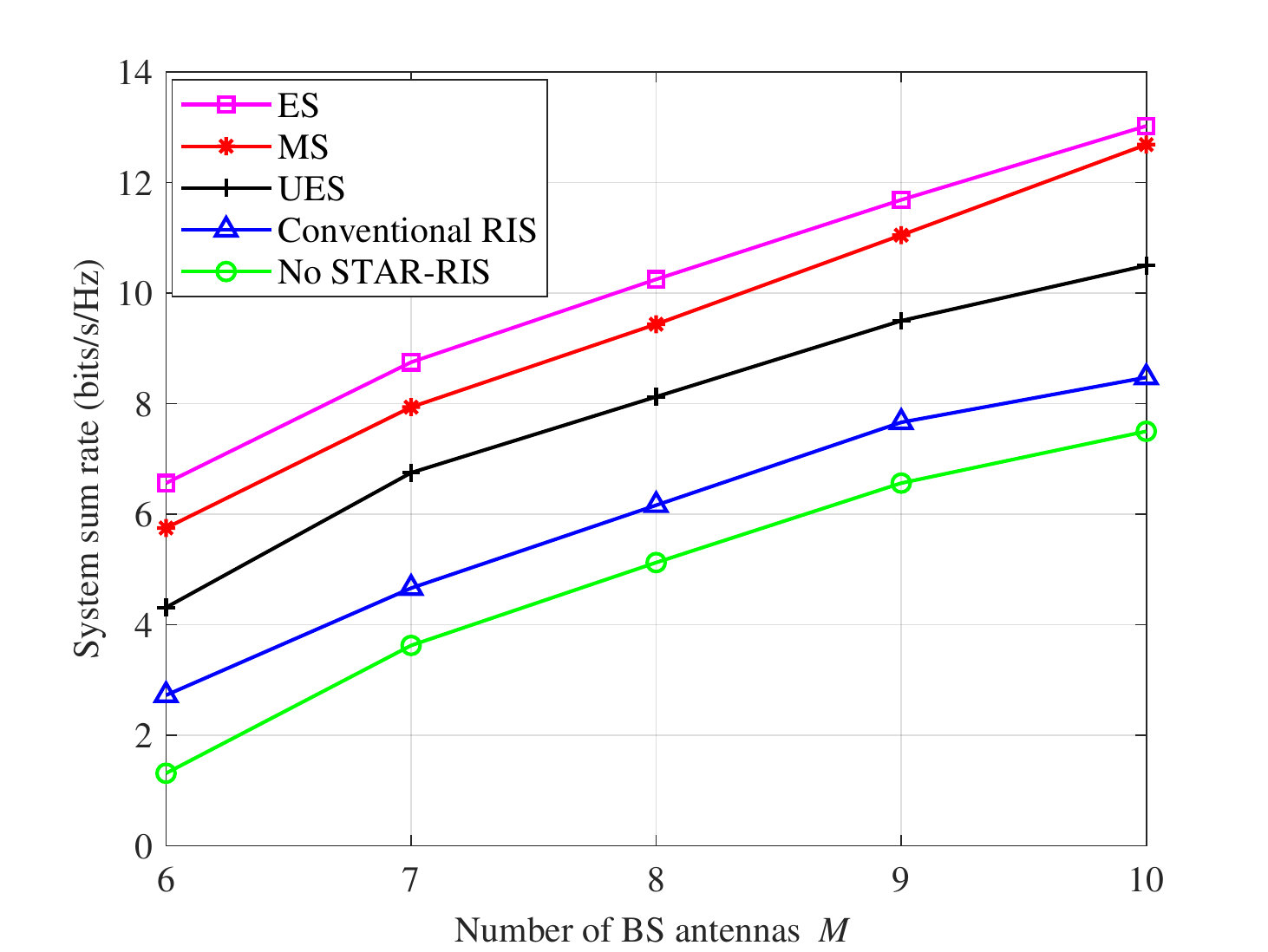}
	\caption{System sum rate versus the number of BS antennas.\label{fig_m}}
\end{figure}
Figure \ref{fig_m} shows the variation of the system sum rate versus the number of BS antennas. From the figure, it can be seen that the system sum rate grows as the number of BS antennas increases. This is because more BS antennas can achieve higher beamforming gain, which improves the performance of the system. At the same time, the ES and MS modes still outperform the other comparison schemes because STAR-RIS deploys a variable number of transmitting and reflecting elements, taking advantage of the degrees of freedom in the system to enhance the signal strength.

\section{Conclusion}\label{sec_con}
In this work, to maximize the system sum rate
under the maximum transmit power constraint of the BS as well as the minimum rate constraint of the users, we jointly designed the BS transmit beamforming and the TARCs matrices of the STAR-RIS for STAR-RIS-UAV aided  CoMP systems. 
Furthermore, in order to effectively solve the nonconvex optimization problem in both protocols, an iterative algorithm based on penalty function was proposed for ES protocol and further extended to MS protocol. 
The simulation results show that STAR-RIS-UAV can significantly improve the system sum rate and compared with the conventional RIS scheme as well as the no-STAR-RIS scheme. More importantly, proposed methods can also be used  when the system serve other UAVs. This conclusion provides useful guidance for the design of STAR-RIS-UAV aided wireless communication systems. 

\ifCLASSOPTIONcaptionsoff
  \newpage
\fi

\bibliographystyle{IEEEtran}
\bibliography{STARRIS_ref}

\end{document}